\documentclass[prd,aps,floatfix,twocolumn]{revtex4-2}
\usepackage{graphicx,graphics,psfrag,amsmath,calc}
\usepackage{epsfig}
\usepackage{color,comment}
\usepackage{amsfonts}
\usepackage{tikz}
\usepackage{hyperref}

\usepackage{capt-of} 
\usepackage[normalem]{ulem} 

\definecolor{vermillion}{rgb}{0.86, 0.18, 0.01}

\newcommand{\utrentoafil}{Physics Department, University of Trento, Via Sommarive 14, I-38123 Trento, Italy}
\newcommand{\tifpaafil}{INFN-TIFPA Trento Institute of Fundamental Physics and Applications, Via Sommarive 14, I-38123 Trento, Italy}
\newcommand{\inqubatorafil}{InQubator for Quantum Simulation (IQuS), Department of Physics, University of Washington, Seattle, WA 98195, USA}
\newcommand{\fnalafil}{Fermi National Accelerator Laboratory, Batavia, IL, 60510, USA}
\newcommand{\lanlafil}{Theoretical Division T-2, Los Alamos National Laboratory, Los Alamos, NM 87545, USA}

\newcommand{\reportnumber}{FERMILAB-PUB-25-0019-ETD,  LA-UR-25-20317}

\usepackage[angle=0,scale=1,color=black,firstpage=true,opacity=1]{background}

\SetBgContents{\small\texttt{\reportnumber}}
\SetBgPosition{current page.north east}
\SetBgHshift{-5.5cm}
\SetBgVshift{-1.5cm}

\begin{document}
\title{Inference of response functions with the help of machine learning algorithms }

\author{
Doga Murat Kurkcuoglu
}
\affiliation{\fnalafil}

\author{
Alessandro Roggero}
\affiliation{\utrentoafil}
\affiliation{\tifpaafil}
\affiliation{\inqubatorafil}

\author{Gabriel~N.~Perdue}
\affiliation{\fnalafil}

\author{Rajan Gupta}
\affiliation{\lanlafil}

%\author{TBD}
\date{\today}

\begin{abstract}
Response functions are a key quantity to describe the near-equilibrium dynamics of strongly-interacting many-body systems.
Recent techniques that attempt to overcome the challenges of calculating 
these \emph{ab initio} have employed expansions in terms of 
orthogonal polynomials.
We employ a neural network prediction algorithm to reconstruct a 
response function $S(\omega)$ defined over a range in frequencies $\omega$.
We represent the calculated response function as a truncated Chebyshev series whose coefficients can be optimized to reduce the representation error. We compare the quality of response functions obtained using coefficients calculated using a neural network (NN) algorithm with those computed using the Gaussian Integral Transform (GIT) method. In the regime where only a small number of terms in the Chebyshev series are retained, we find that the NN scheme outperforms the GIT method.
\end{abstract}

\maketitle

There are a number of problems of scientific interest where quantum computers may play a key role in finding solutions.
In particular, quantum simulations offer a promising set of algorithms for computing quantities of interest that may be beyond the capabilities of classical computers in tractable times.
Indeed, quantum simulation may be one of the first quantum computations to show practical quantum advantage~\cite{Childs_2018}.
Many of the possible use cases in High Energy Physics are discussed in recent reviews~\cite{Klco_2022,Bauer:2022hpo,PRXQuantum.5.037001}. These include simulations of the real time dynamics of quantum systems and alleviating the sign problem limiting classical simulations.
A motivation for this work is that quantum computers may provide valuable contributions to problems relevant to the long-baseline neutrino oscillation experiments \cite{qcforscattering,twopoint}.
One of the prominent examples of this is direct ab-initio calculations of inclusive scattering cross sections on nuclear targets. 

Several schemes for calculating the linear response of strongly correlated many-body systems using quantum computers have been proposed in the recent past~\cite{chiesa2019quantum,roggero_carlson_2019,Ciavarella2020,Tong2021,twopoint,Libbi2022}. An effective use of near term quantum computers for this goal relies on the extraction of suitably defined energy moments which contain the non-perturbative information and many-body observables, like scattering cross sections, can then be reconstructed with the help of classical algorithms~\cite{Somma_2019,Roggero2020_git,PRXQuantum.2.020321,PRXQuantum.3.010318,kiss2024quantum,kiss2024earlyfaulttolerantquantumalgorithms}. This strategy has been employed in various forms in classical algorithms for a long time and remarkable progress has been achieved in recent years using classical many-body techniques like Quantum Monte Carlo~\cite{Lovato2015,Lovato2016,Pastore2020,Lovato_2020,Andreoli_2022} and Coupled Cluster theory~\cite{Bacca2013,Bacca2014,Sobczyk2021,Sobczyk2022,sobczyk202316o,sobczyk202340ca,acharya202416oelectroweakresponsefunctions,sobczyk2024spinresponseneutronmatter}. 
These methods are well suited for calculations of static properties of nuclear many-body systems, however, information about the scattering cross sections, as encoded in the so-called response function $S(\omega)$, are accessed through suitably defined integral transform. 
The dependence on the energy transfer $\omega$ of the scattering cross section is captured by the following response function
\begin{equation}
S(\omega) = \frac{\langle \Psi_0 \lvert \hat{O}^\dagger \delta(\hat{H} - \omega)\hat{O}\rvert\Psi_0\rangle}{\langle \Psi_0 \lvert \hat{O}^\dagger\hat{O} \rvert\Psi_0 \rangle}\;,
\end{equation}
with $\rvert\Psi_0\rangle$ the ground state of the nuclear target, $\hat{H}$ the Hamiltonian describing the target and $\hat{O}$ an excitation operator describing the interaction between the target and the probe. 
%\gabe{We need to define $\hat{H}$.}
With this definition, the integral of $S(\omega)$ over frequency  is equal to one. An integral transform $\Phi(\omega)$ can then be defined as 
\begin{equation}
\begin{split}
    \Phi(\omega) &= \int K(\omega, \nu)S(\nu)d\nu\\
    &= \frac{\langle \Psi_0 \lvert \hat{O}^\dagger K(\omega, \hat{H})\hat{O}\rvert\Psi_0\rangle}{\langle \Psi_0 \lvert \hat{O}^\dagger\hat{O} \rvert\Psi_0 \rangle}\;, 
    \label{eq:phi}
\end{split}
\end{equation}
with the kernel $K(\omega, \nu)$ defining the integral transform.

The main advantage is  $\Phi(\omega)$ can be estimated as a ratio between ground state expectation values provided the, so far arbitrary, kernel $K$ is chosen appropriately.

For numerical schemes based on imaginary-time Monte Carlo sampling, an exponential kernel --- leading to the Laplace transform~\cite{Carlson_1992,Carlson2015} --- is typically employed, while for schemes working in occupation number basis --- like the Coupled Cluster method --- a Lorentzian or Gaussian kernel is  used~~\cite{Efros_1994,Efros_2007,Roggero2020_git}.

A popular numerical technique in condensed matter physics relies instead on integral kernels defined as finite linear combinations of orthogonal polynomials (see e.g.~\cite{KPM2006}).
This approach has also been used recently for nuclear physics applications on both classical~\cite{Sobczyk2022,sobczyk202316o,sobczyk2024spinresponseneutronmatter} and quantum~\cite{Hartse2023,kiss2024quantum} computing platforms, including the possibility of quantifying the errors induced by the integral transform~\cite{Roggero2020_git,Sobczyk2022_git}.
The central goal of this class of methods is to use the computed integral transform $\Phi(\omega)$ to reconstruct the original response function $S(\omega)$.

Unfortunately, this procedure leads to an ill-posed inversion problem that is, in general, extremely sensitive to small errors in the input data~\cite{Gl_ckle_2009,Barnea_2010}.
This issue can be circumvented when the goal is to only estimate some observables connected with integrated properties of the response, e.g., the electric dipole polarizability of nuclei~\cite{Miorelli2016} or the impurity contribution to the thermal conductivity in the outer crust of neutron stars~\cite{Roggero_2016}.
Since these can be obtained directly from the integral transform, one avoids the inversion step.

For the general case, multiple techniques have been developed to perform numerically stable inversions at the price of introducing systematic errors which are not always easy to quantify~\cite{Silver1990,Vitali2010,Burnier2013,Kades2020,Raghavan2021,raghavan2023uncertainty}.
When the kernel satisfies certain regularity conditions it can be shown that stable reconstructions can be performed by limiting the final energy resolution in a way that takes into account the systematic errors~\cite{Roggero2020_git}.
For this purpose, it was found convenient to define integral kernels to be $\Sigma$-\textit{accurate} with \textit{resolution} $\Delta$ if the following holds~\cite{Roggero2020_git}
\begin{equation}
\inf_{\omega_0\in[E_{0},E_{max}]}\int_{\omega_0-\Delta/2}^{\omega_0+\Delta/2} d\nu K(\nu, \omega_0) \geq 1-\Sigma\;.
\label{eq:kernel_definition}
\end{equation}
In this expression the frequency $\omega_0$ over which one is extremizing is taken to be the full spectrum of the Hamiltonian $[E_0,E_{max}]$. The intuition behind this definition is that for a small value of $\Sigma$ and an arbitrary but fixed frequency $\omega_0$, a kernel satisfying Eq.~\eqref{eq:kernel_definition} has most of its support in an interval of size $\Delta$ around the target frequency. One can then expect the integral transform $\Phi(\omega)$ obtained employing such a kernel to be an approximation of the original response $S(\omega)$ smoothed over a frequency range of order $\Delta$. 
This intuition can be made more rigorous and allows one to show that when using these integral kernels, it is possible to produce histograms of $S(\omega)$ with rigorous error bars~\cite{Sobczyk2022_git}.

In cases where either the required energy resolution is small or the preparation of the ground state $\rvert\Psi_0\rangle$ is complicated, simulations on quantum computers could prove useful for carrying out the calculation in an efficient manner, thus opening the possibility of expanding the range of targets where the calculation of scattering cross sections becomes possible (see e.g.~\cite{Klco_2022,Bauer:2022hpo} for recent reviews).

Inspired by recent work employing techniques from Machine Learning(ML) for the problem of inversion of integral transforms in Quantum Monte Carlo simulations~\cite{Kades2020,Raghavan2021,raghavan2023uncertainty}, we here develop a numerical technique that uses Neural Networks to tailor the integral kernel used in the definition of $\Phi(\omega)$ to a specific application by training it over template response function generated from an underlying model. The hope is that using such tailored kernel functions, and for fixed number of moments, a higher energy resolution will be achievable than using general purpose kernels like the Gaussian.

\section{Integral Transform Method}
In this section, we present in more detail the Integral Transform method using Chebyshev polynomials following the presentation in Ref.~\cite{Sobczyk2022_git}. In order to simplify the notation used in the following derivation, it is convenient to introduce a normalized state obtained by perturbing the ground-state with $\hat{O}$ as follows
\begin{equation}
\rvert\psi_O\rangle = \frac{\hat{O}\rvert\Psi_0\rangle}{\sqrt{\langle \Psi_0 \lvert \hat{O}^\dagger\hat{O} \rvert\Psi_0 \rangle}}\;.
\end{equation}

Since we are interested in simulations where the many-body system is defined over a finite basis, the Hamiltonian of the system, $\hat{H}$, can be represented as a finite matrix of dimension $N$. 
In order to avoid complications when using Chebyshev polynomials, the spectrum of the Hamiltonian $[E_0,E_{max}]$ is mapped into the interval $[-1,1]$ by defining a shifted and scaled Hamiltonian 
\begin{equation}
\widetilde{\hat{H}} = \frac{2\hat{H}-(E_0+E_{max})}{E_{max}-E_0}\;.
\end{equation}
In situations when suitably tight bounds for $E_0$ and $E_{max}$ are not available, we can instead use $\widetilde{\hat{H}}=\hat{H}/\Lambda$ with $\Lambda\geq\|\hat{H}\|$ as a computable upper bound on the spectral norm of the Hamiltonian (for applications on quantum computers this is always obtainable in an efficient way).
We will assume from here on that this redefinition of the Hamiltonian of the system has been performed and refer to it, for brevity, by $\hat{H}$.

The discrete spectrum allows us to express the response function as
\begin{equation}
S(\omega)=\sum_{k=1}^N s_k\delta(\lambda_k-\omega)\;,
\label{eq:sofomega}
\end{equation}
where $\lambda_k$ are the $N$ eigenvalues of $\hat{H}$ with eigenvectors $\rvert \Psi_k\rangle$. The probabilities $s_k$ are given by the squared overlaps
\begin{equation}
\label{eq:skdef}
s_k=\left|\langle\Psi_k\vert\psi_O\rangle\right|^2\quad\quad\text{with}\quad\quad\sum_{k=1}^Ns_k=1\;,
\end{equation}
Using this notation, an integral transform of the response function $S(\omega)$ then becomes
\begin{equation}
\begin{split}
\Phi(\omega)&=\langle\psi_O\lvert K(\omega,\hat{H})\rvert\psi_O\rangle=\sum_{k=1}^Ns_k K(\omega,\lambda_k)\;.
\end{split}
\end{equation}

Given our goal to design an integral transform informed by the expected properties of the response function, we aim to determine an integral kernel $K$ such that its corresponding integral transform minimizes the following cost function
\begin{equation}
    C(\Delta) = \max_{\omega_0\in [-1,1]}\left|\int \limits_{\omega_0-\Delta/2}^{\omega_0+\Delta/2}d\omega \left(\Phi(\omega)-S(\omega)\right)\right|\;.
    \label{eq:cost_resolution_for_discussion}
\end{equation}
The intuition behind this definition is the same that led to the introduction of the condition in Eq.~\eqref{eq:kernel_definition} in Ref.~\cite{Roggero2020_git}: when $C(\Delta)$ is small, the  strength of the response function in an interval of size $\Delta$ centered anywhere within the range of the spectrum is well approximated by the strength evaluated using the integral transform $\Phi(\omega)$ instead. Indeed if the kernel is $\Sigma$-\textit{accurate} with \textit{resolution} $\Delta$ we have $C(\Delta)\leq\Sigma$.

In this work we attempt to optimize integral kernels expressed as a finite sum of Chebyshev polynomials $T_i(x)$ as follows
\begin{equation}
\label{eq:kernel}
K_M(\omega,\nu)=\sum_{i,j=0}^{M-1} b_{ij} T_i(\omega)T_j(\nu)\;.
\end{equation}
Note that this is not equivalent to the standard construction used in the Kernel Polynomial Method~\cite{KPM2006} or the Gaussian Integral Transform~\cite{Roggero2020_git,Sobczyk2022_git} since both of these include non polynomial contributions in $\nu$.
We chose this simplified class of kernels because the discussion of the optimization via neural-networks is easier in this case, but the strategy we propose in this work can be generalized easily to account for the differences.
The main advantage of using Chebyshev polynomials to capture the $\nu$ dependence of the kernel is that the integral transform of the response function can be evaluated as
\begin{equation}
\Phi(\omega) = \sum_{i,j=0}^{M-1}b_{ij}T_i(\omega)\langle\psi_O\lvert T_j(\hat{H})\rvert\psi_O\rangle\;,
\label{eq:int_transf}
\end{equation}
where the expectation value appearing on the right hand side is the $j$-th Chebyshev moment of the Hamiltonian evaluated on the state $\rvert\psi_O\rangle$, which we will denote as
\begin{equation}
\label{eq:cheb_mom}
m_j=\langle\psi_O\lvert T_j(\hat{H})\rvert\psi_O\rangle\;.
\end{equation}

The use of Chebyshev polynomials as a basis is particularly convenient since the moments $m_j$ which encode the information about the response can be calculated efficiently on quantum computers~\cite{doi:10.1137/16M1087072,Subramanian_2019,Gily_n_2019,Roggero2020_git} and, at least for some systems, approximated accurately using Coupled Cluster theory\cite{Sobczyk2022,sobczyk202316o,sobczyk2024spinresponseneutronmatter}.

\section{Neural Network }

In this section, we describe the use of neural networks (NN) in this problem. 
NNs are useful in making predictions in physics to find the thermodynamic quantities, or to infer the quantum state based on the circuit training data. 
Here, we use the NN to infer the response signal from a few given moments,  $m_j$. 
In order to compute the $m_j$ moments from Eq.~\eqref{eq:cheb_mom}, only the $|\psi_O\rangle$ state, and the Hamiltonian $\hat{H}$ are needed. 
The elements $b_{ij}$ of the coefficient matrix defining the integral kernel in Eq.~\eqref{eq:kernel} are left as unknown parameters that will be optimized and predicted by the neural network. 

Ideally, the optimization procedure would work as follows. First, generate a set of response functions $S^n(\omega)$ that are used for training, validation and testing. For each $S^n(\omega)$ in the set, compute the corresponding Chebyshev moments $\vec{m}^n$. Then choose a target resolution $\Delta_T$ and a budget of moments $M$ and determine the optimal response function as a map 
\begin{equation}
\vec{m} \quad\longrightarrow\quad b^{opt}_{ij}(\vec{m})
\end{equation}
depending on an input set of $M$ Chebyshev moments $\vec{m}$. The goal being to create a multilayer NN, train it with $m_j$ moments and infer the $b_{ij}$ matrix from it.

\begin{figure}
\includegraphics[width=0.9\linewidth]{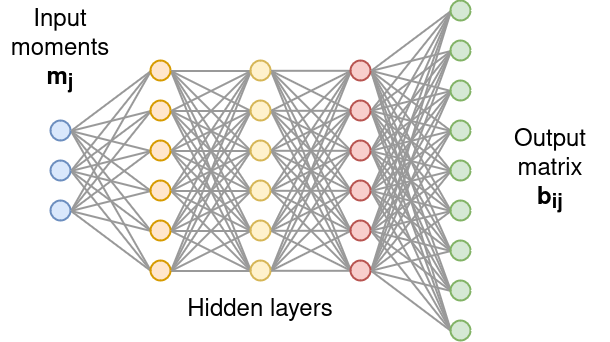} 
\caption{A schematic of our inference neural network algorithm. The inputs are the $M$ moments $m_j$ and the output is the $M^2$ component $b_{ij}$ vector.  }
\label{fig:neuralnetwork_scheme}
\end{figure}

In this work we represent this map using a multilayer perceptron with three hidden layers as depicted in the diagram in Fig.~\ref{fig:neuralnetwork_scheme} and described in Table~\ref{tab:nn_params}.
In all, the NN has a three hidden, one input and one output MLP layers. The input layer with $M$ nodes normalizes the $m_j$ moments. The three hidden layers use ReLU activation functions and have 30, 25 and 40 nodes, respectively. We use Adam with a variable learning rate to minimize the cost function of the NN.

In principle, the NN architecture may be optimized, but for initial demonstration purposes an easy-to-understand design is desirable.
Similarly, hyper-parameter optimization would likely find different preferred values than the ones presented in Table~\ref{tab:nn_params}.
However, in a given application, users will likely be sensitive to data generation costs and processes, and those features will likely differ from this case. For demonstration purposes, we chose one set of parameters based on intuition and experience. Very little tuning was done as the goal is not to surpass a state-of-the-art benchmark, but rather to establish that the technique functions as hypothesized.\looseness-1

\begin{table}[t]
\begin{tabular}{c|c}
$L_1$ & 30 \\\hline
$L_2$ & 25 \\\hline
$L_3$ & 40 \\\hline
\text{learning rate} & 0.00025 \\\hline
\text{batch size} & 80 \\\hline
\text{epochs} & 300 \\\hline
\text{number of data for training} & 2000 \\\hline
\text{number of data for validation} & 400 \\\hline
\text{number of data for test} & 100 \\\hline
\end{tabular}
\caption{Parameters of the NN used in this work. $L_1$, $L_2$ and $L_3$ are the number of hidden nodes in each layer. We used the Adam optimizer, choose a small learning rate and a large number of epochs to have smooth controlled convergence. The split of the 2500 data points for each $(M,\Delta)$ is given in the last three lines.\label{tab:nn_params}}
\end{table}

The cost function $C(\Delta)$ in Eq.~\eqref{eq:cost_resolution_for_discussion} is rather demanding to evaluate precisely. In order to approximately minimize the cost function over $\omega_0$ we can calculate this quantity on a uniform grid of points with spacing $d<\Delta$. For simplicity, we pick $K$ points so that the lattice spacing in the energy interval $[-1,1]$ is $d=2/K$, and we choose $\Delta=2L/K$ for some integer $1\leq L\ll K$. This means that an interval of size $\Delta$ is given by points $L$ units apart and the cost function can be then approximated as
\begin{equation}
    C(\Delta) = \max_{l=0,\dots,K}\left|\int \limits_{-1+ld}^{-1+(l+L)d}d\omega \left(\Phi(\omega)-S(\omega)\right)\right|\;.
\label{eq:discrete_cost}
\end{equation}
The integral inside the absolute value can then be done exactly integrating separately the response function $S(\omega)$ from Eq.~\eqref{eq:sofomega} and the Chebyshev moments $T_i(\omega)$ in the expansion of the integral transform Eq.~\eqref{eq:int_transf}.
During the NN training we opted to use instead an upperbound $C_U(\Delta)$, which can be evaluated more efficiently and is also smoother. In order to obtain $C_U(\Delta)$ we first notice that the cost function in Eq.~\eqref{eq:discrete_cost} can be expressed as $C(\Delta)=\|\vec{\mathcal{C}}\|_\infty$ where $\|\cdot{}\|_\infty$ denotes the infinity vector norm and
\begin{equation}
\mathcal{C}_l = \int \limits_{-1+ld}^{-1+(l+L)d}d\omega \left(\Phi(\omega)-S(\omega)\right)\;.
\end{equation}
We can then introduce $C_U(\Delta)$ as
\begin{equation}
C_U(\Delta)= \|\vec{\mathcal{C}}\|_2\geq \|\vec{\mathcal{C}}\|_\infty=C(\Delta)\;,
\label{eq:actual_cost_function}
\end{equation}
where $\|\cdot{}\|_2$ is the standard vector 2 norm. The main advantage of this change in cost function is that $C^2_U(\Delta)$ is a quadratic function of the coefficient matrix $b_{ij}$ and can be minimized straightforwardly using the least squares method. The resulting optimal matrix $b^{(min)}_{ij}$ is used as the target for the NN training with a mean square error (MSE) cost function.

\section{Response Functions}

For the numerical calculations carried out in this work, the dimension of the Hamiltonian matrix is fixed to $N=200$. These 200 energy eigenvalues, $\lambda_k$, with $k=1,\dots,N$, are chosen  using a uniform random sampling in the full energy interval normalized to $[-1,1]$. For each $N$ eigenvalues $\{\lambda_k\}$,  $N_S=2500$ data points, labeled $\lambda^n_k$ with $n=1,\dots,N_S$ are generated as described below.  

Similarly, for each set of eigenvalues, 
$\{\lambda_k\}$, we generate a response function
\begin{equation}
\label{eq:model_response}
S^n(\omega) = \sum_{k=1}^N s^n_k\delta(\lambda_k^n-\omega)\;,
\end{equation}
by computing the probabilities $s^n_k$ according to a distribution that resembles the typical features of the physical response function we are interested in. For applications of this scheme to study the nuclear response in the quasi-elastic regime, which is particularly important for planned and ongoing neutrino-nucleus experiments, we use a model based on a skewed Gaussian function
\begin{equation}
\label{eq:skgauss}
S_S(\omega;\mu,\sigma,\alpha,\Gamma) = \Gamma e^{-\frac{(\omega-\mu)^2}{2\sigma^2}}\left[1+ \mathrm{erf}\left(\frac{\omega\alpha }{\sigma\sqrt{2}}\right)\right]\;,
\end{equation}
depending on 4 parameters: a location $\mu$, a width $\sigma$, a skewedness parameter $\alpha$ and a normalization constant $\Gamma$ that is determined by satisfying \eqref{eq:sknorm}. 

The 2500 data points, and consequently the individual response functions $S^n(\omega)$, are generated by choosing fifty values each for $\alpha$ and $\sigma$. The fifty $\alpha$ are chosen uniformly within the range $\alpha\in[0,2.5]$ to mimic a stronger high energy tail in the response. The fifty $\sigma$  are chosen uniformly within   $\sigma\in[\sigma_{\min},0.4]$ for a suitable $\sigma_{\min}$ that is chosen to be $\sigma_{\min}=\Delta$. This choice is motivated by the fact that for signals with $\sigma<\Delta$, a good approximation of the response function that minimizes our cost $C(\Delta)$ can be obtained using only the first two moments $m_0$ and $m_1$ leading to an undetermined optimization problem.
For each of these 2500 data points, $\mu$ is chosen randomly within the interval $\mu \in[-0.3,0.3]$ ensuring most of the strength is in the central region of the energy range.

 For each response function $S^n$, the individual probabilities $s_k^n$ are constructed as 
\begin{equation}
\widetilde{s_k^n}=S_S(\lambda_k^n;\mu_n,\sigma_n,\alpha_n,1)\;\Rightarrow\;s_k^n=\frac{\widetilde{s_k^n}}{\sum_k\widetilde{s_k^n}}\;,
\label{eq:sknorm}
\end{equation}
in order to ensure the normalization condition in Eq.~\eqref{eq:skdef}.

The corresponding integral transform is then given by 
\begin{equation}
\Phi^n_M(\omega;b_{ij}) = \sum_{i,j=0}^M b_{ij}T_i(\omega)m^n_j\;,
\end{equation}
where the $m_j^n$ moments are calculated independently for each 
$S^n(\omega)$. This gives a total of $M \times N_S$ numbers. 
In other words, for each of the $N_S$ points, the $M$ moments, $m_j$, are calculated directly from the model response function $S^n$ as follows:
\begin{equation}
m_j = \int_{-1}^1  \frac{d\omega}{\sqrt{1-\omega^2}} S^n(\omega) T_j(\omega)\;.
\end{equation}
This integral becomes a sum when Eq.~\eqref{eq:model_response} is used. 
 The calculation and comparison of the cost function using the trained NN and the GIT~\cite{Roggero2020_git,Sobczyk2022_git} algorithm starts 
with these $M \times N_S$ numbers.

\section{Numerical Results}

\begin{figure*}[ht]
\includegraphics[width=\linewidth]{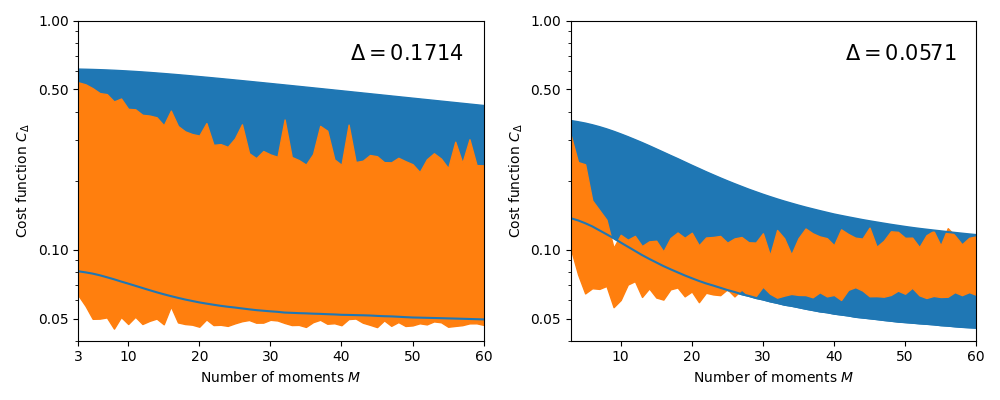} 
\caption{(Color online) The cost function vs. the number of moments $M$ for two values of resolution: $\Delta=0.1714$ (left panel) and $\Delta=0.0571$ (right panel). The orange band shows the 90\% confidence interval of results obtained using the trained NN  while the blue band corresponds to the same confidence interval over the same 100 test data points but for results obtained using the GIT method~\cite{Roggero2020_git,Sobczyk2022_git}.}
\label{fig:costofM}
\end{figure*}

The choice $N_s=2500$ data points   for each fixed  ($M, \Delta$), selected as explained above, was a reasonable match to our NN's capacity. This dataset was split into 2000 training, 400 validation, and 100 test points, and the NN trained using the hyperparameters presented in Table \ref{tab:nn_params}. The  response signal is inferred from the training  by minimizing $C_U(\Delta)$ defined in Eq.~\eqref{eq:actual_cost_function}with $K=70$.

The results for the cost function $C_\Delta$ obtained with the integral kernel expressed in terms of the NN are compared with the general purpose GIT method based on a Gaussian kernel~\cite{Roggero2020_git,Sobczyk2022_git}. The truncated kernel that approximates a Gaussian with variance $\lambda$ is expressed in terms of Chebyshev polynomials as
\begin{equation}
K^{(G)}_M(\omega,\nu;\lambda) = \sum_{i=0}^{M-1} c^{[M,M]}_{i}(\nu;\lambda)T_i(\omega)
\end{equation}
with coefficient functions $c^{[M,M]}_{i}(\nu;\lambda)$ chosen to provide a good representation of a Guassian over $M$ moments and given explicitly in Eq.(A20) of Ref.~\cite{Sobczyk2022_git}. For a given choice of resolution $\Delta$ and number of moments $M$, we obtain the value of $\lambda$ to minimize the approximation error made when  the response function is replaced with an integral transform in Eq.~\eqref{eq:kernel_definition} 
where   $\Sigma$ accounts for the contribution of the tails (for more details see Ref.~\cite{Sobczyk2022_git}).

Figure~\ref{fig:costofM} shows a comparison of the results obtained with the GIT method and the new NN-based technique. The two panels show the 90\% confidence intervals for the value of the cost function, Eq.~\eqref{eq:actual_cost_function}, obtained using the 100 data points reserved for testing, as a function of the number of moments $M$ used in the reconstruction. The left panel corresponds to a wider resolution $\Delta=12/70\approx0.1714$ while the right panel shows results for a narrower resolution $\Delta=4/70$. As $\Delta$ is decreased, the GIT method provides increasingly better results when a large number of moments are included, i.e., the right panel shows this happens for $M > 60$ for $\Delta=4/70$. On the other hand, the new NN approach shows saturation early, that will eventually be outperformed by the GIT method at large values of $M$. To summarize our main result: for narrower $\Delta$, as shown for $\Delta=4/70$, the output of the 
NN method saturates for $M \approx 10$. Thus, when doing calculations keeping a small number of moments, $M\lesssim15-20$, the NN method is able to provide a good integral transform much faster. We can interpret the result as a consequence of the additional information employed in the training of the network which is not used for a general purpose kernel like the Gaussian used in GIT.

\section{Conclusions}

Several state of the art numerical techniques for studying the linear response of strongly correlated systems in nuclear physics rely on the physical information encoded in energy moments of a perturbed initial state. In both classical and quantum algorithms, the use of Chebyshev polynomials for the definition of these energy weights, and the subsequent reconstruction of the nuclear response function, has been proposed as an effective and controllable strategy. In this work we explored the use of Neural Networks to increase the allowed energy resolution when working with a limited number of input moments by relying on the availability of a suitable training set of response functions that can be motivated by the physics of the problem. Employing this physics inspired training, we were able to show that when a limited amount of information (in the form of Chebyshev moments) is available, NN-based strategies can outperform more general reconstruction schemes (like the GIT) that are more agnostic in design and do not directly exploit the features present in physically realistic situations.

\subsection{Future work}

The goal of this work was not to produce the most optimal ML model or to outperform a published state of the art (SOTA) result.
Instead, our primary focus was on demonstrating the technique works under at least some set of reasonable conditions, and from that perspective, using a simple and easy-to-understand ML algorithm was suitable. We, however, outline a number of improvements that should be considered when deploying this technique in practice, and for future research.

First, it would be appropriate to use a process like neural architecture search (see \cite{white2023neuralarchitecturesearchinsights} for a recent review) to optimize the number and size of the hidden layers, or to consider the use of specialized layers.
Furthermore, there are multiple hyperparameters (e.g., learning rate, learning rate schedule, batch size, etc.) that should be tuned using an efficient search strategy (see, e.g. \cite{deephyper} and similar codes).

The next obvious improvement would be to enlarge the training dataset (and increase model capacity accordingly).
Our dataset was chosen for ease of generation on laptop-scale resources, but a modest investment of time would parallelize the code for dataset generation and  the analysis, and allow the utilization of high performance computing (HPC) resources to dramatically increase the dataset size while still getting results in very little ``wall-clock'' time.

Finally, there are interesting opportunities for architectural experimentation.
For example, one could deploy an auto-encoder structure to find an encoding into a fixed latent dimensionality that could improve feature selection for a variety of tasks. 
We could also study linear combinations of moments, and study symmetries in the dataset and try to engineer network layers built to keep those symmetries invariant (a so-called ``physics-inspired'' network).

\section*{Acknowledgements}

We thank Joseph Carlson for many discussions. This document was prepared using the resources of the Fermi National Accelerator Laboratory (Fermilab), a U.S. Department of Energy (DOE), Office of Science, HEP User Facility. 
Fermilab is managed by Fermi Research Alliance, LLC (FRA), acting under Contract No. DE-AC02-07CH11359.
This work was supported by the U.S. Department of Energy, Office of Science, National Quantum Information Science Research Centers, Quantum Science Center.
This manuscript has been authored by Fermi Forward Discovery Group, LLC under Contract No. 89243024CSC000002 with the U.S. Department of Energy, Office of Science, Office of High Energy Physics.

D.M.K. and G.N.P. were supported for this work by the DOE/HEP QuantISED program grant ``HEP Machine Learning and Optimization Go Quantum,'' identification number 0000240323. R.G. was supported by the U.S. Department of Energy, Office 
of Science, High Energy Physics, QuantiSED Contract KA2401032 and AI/ML for HEP 
contract KA2401045 to Los Alamos National Laboratory.

\bibliography{references.bib}

\end{document}